\documentclass{article}
\usepackage{spconf,amsmath,graphicx}
\usepackage{amssymb}
\usepackage{booktabs}
\usepackage{multirow}
\usepackage{pifont}
\usepackage{sidecap}
\usepackage{siunitx}
\usepackage{subcaption}
\usepackage{xcolor}
\usepackage{xparse}

\usepackage[hidelinks]{hyperref}
\hypersetup{
    colorlinks=true,
    linkcolor=red,
    citecolor=green,
    filecolor=magenta,      
    urlcolor=blue
    }

\def\x{{\mathbf x}}

\renewcommand{\and}{\qquad}

\title{Automatic Music Sample Identification with Multi-Track\\ Contrastive Learning}
\name{Alain Riou$^1$ \and Joan Serrà$^1$ \and Yuki Mitsufuji$^{1,2}$}
\address{$^1$ Sony AI \qquad $^2$ Sony Group Corporation}
\newcommand{\CC}{\mathbb{C}}
\newcommand{\RR}{\mathbb{R}}

\renewcommand{\x}{\mathbf{x}}
\newcommand{\X}{\mathbf{X}}
\newcommand{\y}{\mathbf{y}}
\newcommand{\z}{\mathbf{z}}
\newcommand{\yt}{\bar{\y}}

\newcommand{\FF}{\mathcal{F}}
\newcommand{\LL}{\mathcal{L}}

\newcommand{\yart}{\bar{\y}_{\text{art}}}
\newcommand{\yref}{\bar{\y}_{\text{ref}}}
\newcommand{\zart}{\z_{\text{art}}}
\newcommand{\zref}{\z_{\text{ref}}}

\newcommand{\qq}{\mathbf{q}}
\newcommand{\rr}{\mathbf{r}}

\newcommand{\citep}[1]{\cite{#1}}
\newcommand{\citet}[1]{\cite{#1}}

\ExplSyntaxOn
\NewDocumentCommand{\dropfirst}{m}
 {
   \tl_tail:n {#1}
 }
\ExplSyntaxOff
\newcommand{\ci}[1]{\color{gray}\scriptsize $\pm$ \dropfirst{#1}}

\begin{document}
\ninept
\maketitle
\begin{abstract}
Sampling, the technique of reusing pieces of existing audio tracks to create new music content, is a very common practice in modern music production.
In this paper, we tackle the challenging task of automatic sample identification, that is, detecting such sampled content and retrieving the material from which it originates.
To do so, we adopt a self-supervised learning approach that leverages a multi-track dataset to create positive pairs of artificial mixes, and design a novel contrastive learning objective. We show that such method significantly outperforms previous state-of-the-art baselines, that is robust to various genres, and that scales well when increasing the number of noise songs in the reference database. In addition, we extensively analyze the contribution of the different components of our training pipeline and highlight, in particular, the need for high-quality separated stems for this task.
\end{abstract}
\begin{keywords}
Sample identification, contrastive learning, multi-track, self-supervised learning, artificial mixes
\end{keywords}
\section{Introduction}
\label{sec:intro}

\begin{figure*}
    \centering
    \begin{subfigure}{0.48\textwidth}
        \includegraphics[width=\linewidth]{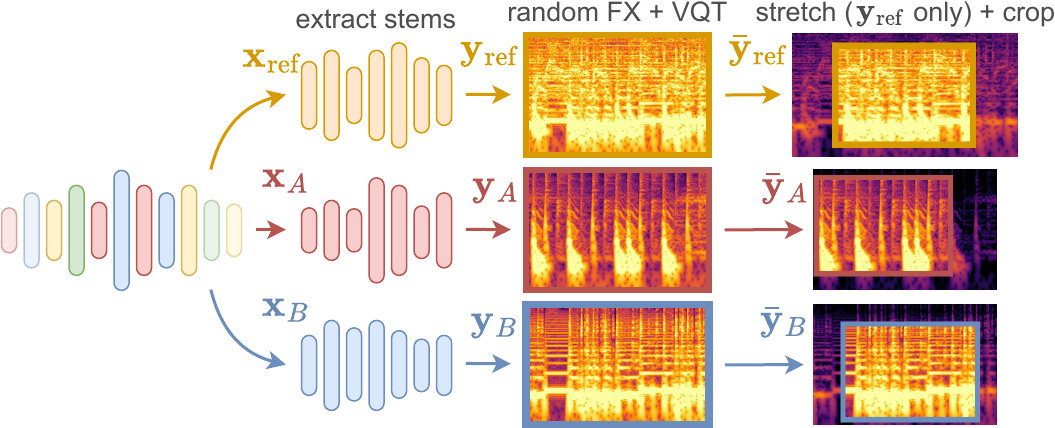}
        \caption{Extraction of audio chunks, transforms, and VQT}
        \label{fig:main-chunks}
    \end{subfigure}
    \hfill
    \begin{subfigure}{0.3\textwidth}
        \includegraphics[width=\linewidth]{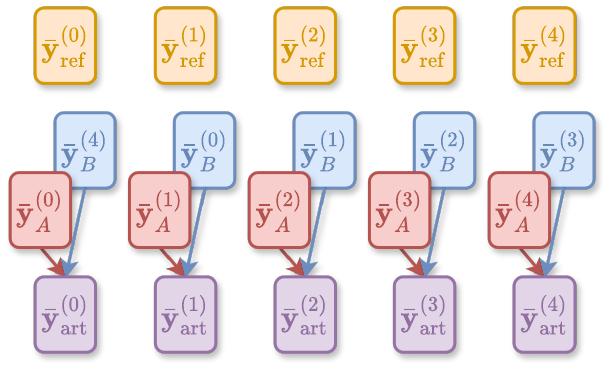}
        \caption{Creation of positive pairs $(\yref, \yart)$}
        \label{fig:main-pairs}
    \end{subfigure}
    \hfill
    \begin{subfigure}{0.205\textwidth}
        \includegraphics[width=\linewidth]{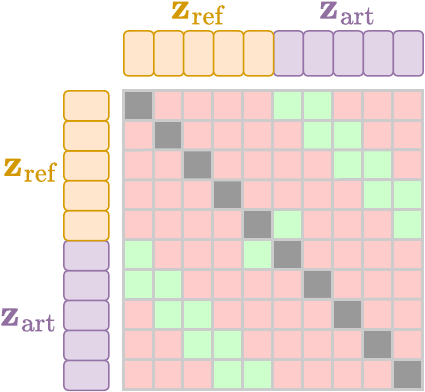}
        \caption{Matrix of our contrastive loss}
        \label{fig:main-loss}
    \end{subfigure}
    \label{fig:main}
    \caption{
        Overview of our training pipeline.
        \textbf{(a)} We crop a random chunk from a multi-track audio and use the separated stems to create $\x_\text{ref}$, $\x_A$, and $\x_B$. We then apply audio effects, convert them to the \emph{complex} VQT domain, time-stretch $\y_{\text{ref}}$ and crop the resulting VQT to obtain $\yref$, $\yt_A$ and $\yt_B$.
        \textbf{(b)} Given a batch of $N$ elements ($N = 5$ here), we construct the $\yart$ by applying a circular shift to the $\yt_B$ and adding them to the $\yt_A$.
        \textbf{(c)} After passing $\yref$ and $\yart$ through the neural network $\FF$ to obtain $\zref$ and $\zart$, we construct the similarity matrix $\sigma \in \RR^{2N \times 2N}_{\geq 0}$ and optimize our contrastive loss $\LL$. Green, red and gray squares denote positive, negative and ignored pairs, respectively.
    }
\end{figure*}

Sampling is a widely used technique that involves reusing existing sound material to create new musical tracks. The sampled material is typically transformed through operations such as pitch-shifting or time-stretching, which makes this practice truly artistic and, sometimes, even political~\citep{SamplingPolitics}. Sampling's origins go back to the 1940s with \emph{musique concrète}, and it has become a core component of hip-hop production~\citep{Davies1996}. Sampling also plays a major role in electronic music and many other styles~\citep{Park2025}, and has been further democratized nowadays thanks to increasing access to digital music, online sample platforms~\citep{Splice} and, more recently, source separation plugins~\citep{Moises}.

Automatic sample identification has, therefore, emerged as an important task, particularly for intellectual property attribution. However, it remains very challenging: systems must be robust to sound effects and transformations applied by the artists while avoiding false negatives. In addition, they have to operate at relatively small time scales, as just one or a few seconds of audio may be sampled in a new production.

The task of sample identification is strongly related to audio fingerprinting~\citep{Cano2005} and, actually, Van Balen et al.\ pioneered this task in 2013 by building upon a spectral peak-based audio fingerprinting algorithm~\citep{van_balen_sample_2013}. However, unlike audio fingerprinting, where the query and reference are identical up to some recording conditions, here not all stems need to be identical between the query and reference segments.
Another related task is musical version identification~\citep{Yesiler2021}, where one also has to deal with many variations in the matching parts. However, the temporal scope of version identification is usually in the order of tens of seconds~\citep{CLEWS}, much longer than the short audios used in current sampling practice. 
Contrary to audio fingerprinting and version identification, where deep neural networks can be trained on large-scale audio catalogs~\citep{DiscogsVI}, sample identification faces a key practical obstacle: the lack of suitable training data. Although collaborative resources such as WhoSampled.com~\citep{WhoSampled} exist, no large-scale annotated dataset is currently available for training sample identification systems. Moreover, scalability remains a major concern, as sample identification requires searching through massive audio databases at a few-second granularity, where exhaustive one-to-one matching quickly becomes infeasible.

To address the aforementioned challenges, recent methods have turned to self-supervised learning, aiming to embed large audio datasets into latent spaces where the cosine similarity between segments reflects the likelihood of one being sampled from the other. In particular, Cheston et al.~\citep{Cheston2025} propose to build an artificial dataset of sample/reference pairs from non-commercial music databases by leveraging source separation and audio effects. Then, they train a ResNet-IBN~\citep{ResNetIBN} in a self-supervised way by optimizing a multi-task triplet and classification objective on these pairs.
In a subsequent work, Bhattacharjee et al.~\citep{Bhattacharjee2025} simplify the framework by applying audio effects on-the-fly and replacing the multi-task objective from~\citet{Cheston2025} by canonical contrastive learning.
More importantly, they show that the performance of the ResNet-IBN can be reached by a much more lightweight Graph Neural Network.

An important point is that, in such works, positive pairs are created from \emph{single} songs, so that the sample is only a subset (up to some audio transforms) of the original track, whereas in real music production the sampled content is incorporated within the final mix.
Hence, we propose to mitigate this issue by creating positive pairs of artificial mixes using tracks from \emph{different} songs on-the-fly during training. More precisely, our contributions are the following:
\begin{itemize}
    \setlength\leftskip{-1.8em}
    \item We propose a simple way to artificially create positive pairs on-the-fly, and introduce a modified contrastive loss for training our model on a large dataset of multi-track recordings.
    \item We evaluate our model both on the standard hip-hop benchmark and on a private dataset spanning more diverse genres. Notably, we outperform the state-of-the-art by a significant margin, with a +15\,\% boost in mean average precision.
    \item We study the impact of our design choices and of our multi-track training set, and measure the robustness of our model to different hop sizes and a high number of noise songs during evaluation.
\end{itemize}
We release our full training code and pretrained models.\footnote{\url{https://github.com/sony/sampleid}}

\section{Method}

Similar to \cite{Bhattacharjee2025}, we train our system using contrastive learning~\cite{SimCLR} with positive pairs built using the different sources of music recordings. Our work, however, fundamentally differs in the way we build these pairs and the contrastive loss we employ.

\subsection{Extraction of chunks from multi-track recordings}

The procedure for extracting sub-mixes from a multi-track recording is depicted in Fig.~\ref{fig:main-chunks}.
Given a multi-track recording composed of $S$ sources with sample rate $f_s$, we first extract from it an $S$-source random chunk of duration $d$, $\X = (\x_0, \dots, \x_{S-1})^\top \in \RR^{S \times \lfloor d f_s\rfloor}$. We then sample two disjoint subsets $A, B \subset \{0, \dots, S-1\}$. Let $\x_A = \sum_{i \in A} \x_i$, $\x_B = \sum_{i \in B} \x_i$, and $\x_{\text{ref}} = \sum_{i=0}^{S-1} \x_i$. We next randomly apply transforms such as gain and equalizer to all of them (the complete list of transformations is provided in Sec.~\ref{sec:details}).

The resulting chunks are then converted to the time-frequency domain using the Variable-Q Transform (VQT)~\citep{VQT}. We note $\y_A \in \CC^{qb \times w}$ (resp.\ $\y_B$, $\y_{\text{ref}}$) the VQT of as the transformed $\x_A$ (resp.\ $\x_B$, $\x_{\text{ref}}$), where $q$ is the number of octaves spanned by the VQT, $b$ is the number of bins per octave, and $w$ is the number of time frames. 
Next, we apply some augmentation operations to the VQT representation (Fig.~\ref{fig:main-chunks}). For $\y_{\text{ref}}$, we remove the upper and lower half-octaves, we apply a random time-stretch using linear interpolation\footnote{In our early experiments, linear interpolation yielded the same performance as audio-domain time stretch, while being considerably faster.}, and temporally crop with a random offset to obtain $\yt_{\text{ref}} \in \CC^{(q-1)b \times w'}$. For $\y_A$ and $\y_B$, we randomly crop both in the frequency and temporal axes, yielding $\yt_A, \yt_B \in \CC^{(q-1)b \times w'}$. 
These stretch-and-crop operations replace costly time-stretch and pitch-shift manipulations in the audio domain, which are essential for robustness (see Table~\ref{tab:baselines}), but computationally expensive.
Due to the log-frequency scale of the VQT, our random cropping is equivalent to a pitch shift of up to $\pm 6$~semitones. For the time shift, the range is determined by two hyperparameters: the chunk duration $d$ and the crop length $w'$.

\subsection{Creation of positive pairs}

As in SimCLR~\citep{SimCLR}, we pick positive and negative samples from batches during training. However, in contrast to SimCLR, we have two positive pairs per item in the batch. We note $N$ as the batch size. Let $i \in \{0, \dots, N-1\}$.
We note $( \yt_A^{(i)}, \yt_B^{(i)}, \yt_{\text{ref}}^{(i)})$ as the $i$-th element from the batch.
We create \emph{artificial} samples $\yart$ as
\begin{equation}
    \label{eq:positives}
    \yart^{(i)} = \yt_A^{(i)} + \yt_B^{(i - 1) \bmod N}.
\end{equation}
Notice that, by construction, the positive pairs (i.e., the pairs of elements that share some sources in common) are both $(\yref^{(i)}, \yart^{(i)})$ and $(\yref^{(i)}, \yart^{(i+1) \bmod N})$. The process is illustrated in Fig.~\ref{fig:main-pairs}.

\subsection{Contrastive training}

All $\yart$ and $\yref$ representations are embedded on the hypersphere $\mathcal{S}^{m-1}$ by a neural network $\mathcal{F}$, whose parameters are optimized by minimizing the contrastive loss depicted in Fig.~\ref{fig:main-loss}. More precisely, we construct the symmetric matrix $\sigma \in \RR^{2N \times 2N}_{\geq 0}$ such that, for all $i,j \in \{0, \dots, N-1\}$, we have $\zref^{(i)} = \FF(\yref^{(i)})$, $\zart^{(i)} = \FF(\yart^{(i)})$, and
$$\begin{alignedat}{2}
    \sigma_{i,j} &= \exp \!\left( \frac{\zref^{(i)} \cdot \zref^{(j)}}{\tau} \right) ,
    &\quad \sigma_{i,N+j} &= \exp \!\left( \frac{\zref^{(i)} \cdot \zart^{(j)}}{\tau} \right) ,\\
    \sigma_{N+i,j} &= \exp \!\left( \frac{\zart^{(i)} \cdot \zref^{(j)}}{\tau} \right) ,
    &\quad \sigma_{N+i,N+j} &= \exp \!\left( \frac{\zart^{(i)} \cdot \zart^{(j)}}{\tau} \right) .
\end{alignedat}$$
The loss is then expressed as $\LL = \frac{1}{4N} \sum_{i=0}^{2N-1} \LL_i$, with
\begin{equation}
    \label{eq:loss}
    \LL_i =
    \begin{cases}
        - \log \left( \dfrac{ \sigma_{i,i+N} }{\sum_{j \neq i} \sigma_{i,j}} \right) - \log \left( \dfrac{ \sigma_{i,i+N+1} }{\sum_{j \neq i} \sigma_{i,j}} \right) \text{ if } i < N, \\
        - \log \left( \dfrac{ \sigma_{i,i-N} }{\sum_{j \neq i} \sigma_{i,j}} \right) - \log \left( \dfrac{ \sigma_{i,i-N-1} }{\sum_{j \neq i} \sigma_{i,j}} \right) \text{ if } i \geq N,
    \end{cases}
\end{equation}
and the convention that $\sigma_{N-1,2N} = \sigma_{N,-1} = \sigma_{N-1,N} = \sigma_{N,N-1}$.

\begin{table*}[]
    \centering
    \resizebox{0.96\textwidth}{!}{%
    \begin{tabular}{lccccccccc}
        \toprule
         & \multicolumn{4}{c}{Sample100~\citep{van_balen_sample_2013}} &  & \multicolumn{4}{c}{SamplePairs} \\
        \cmidrule{2-5} \cmidrule{7-10}
        Model & mAP ($\uparrow$) & HR@1 ($\uparrow$) &
        mNR ($\downarrow$) & medNR ($\downarrow$) &  & mAP ($\uparrow$) & HR@1 ($\uparrow$) & mNR ($\downarrow$) & medNR ($\downarrow$) \\
        \midrule
        Ours &
        \textbf{0.603}\ci{0.098} & \textbf{0.587}\ci{0.111} &
        \textbf{0.074}\ci{0.036} & 0.003 &  & 
        \textbf{0.450}\ci{0.095} & \textbf{0.430}\ci{0.097} &
        \textbf{0.052}\ci{0.024} & \textbf{0.003} \\
        \quad no time-stretch &
        0.463\ci{0.100} & 0.427\ci{0.112} &
        0.135\ci{0.049} & 0.013 &  & 
        0.301\ci{0.086} & 0.270\ci{0.087} &
        0.101\ci{0.174} & 0.018 \\
        \quad no time-shift &
        0.598\ci{0.100} & 0.573\ci{0.112} &
        0.098\ci{0.047} & \textbf{0.001} &  &
        0.376\ci{0.091} & 0.350\ci{0.093} &
        0.056\ci{0.022} & 0.006 \\
        \quad no pitch-shift &
        0.422\ci{0.100} & 0.413\ci{0.094} &
        0.172\ci{0.060} & 0.033 &  &
        0.355\ci{0.092} & 0.340\ci{0.093} &
        0.171\ci{0.051} & 0.032 \\
        Contrastive baseline &
        0.551\ci{0.101} & 0.533\ci{0.113} &
        0.101\ci{0.041} & 0.003 &  & 
        0.409\ci{0.092} & 0.380\ci{0.095} &
        0.075\ci{0.033} & \textbf{0.003} \\
        \bottomrule
    \end{tabular}}
    \caption{
        Performances of our model on the two considered datasets, with $ \pm$ confidence interval at 95\%.
    }
    \label{tab:baselines}
\end{table*}

\subsection{Training details}
\label{sec:details}

We train our model on a proprietary music dataset consisting of 21\,k~multi-track recordings with a total duration of 1350\,h. The dataset features a diversity of music genres (e.g.,~pop/rock, R\&B, electro, country), with approximately 3\,\% of it corresponding to hip-hop songs. 
We randomly extract audio chunks of $d = \SI{7.2}{\second}$, apply random equalization, gain, and compression with the same hyper-parameters as \citet{Cheston2025}, and convert them to the VQT domain spanning $q = 8$ octaves with $b = 36$~bins per octave, and a hop size of \SI{25}{\milli\second}. This differs from most related works that use 12\,bins per octave. Contrary to the CQT~\cite{CQT}, the VQT introduces a bandwidth parameter $\gamma$ that increases the time resolution in low frequencies~\cite{VQT}. We choose $\gamma = 7$ and study the influence of the frontend in Sec.~\ref{sec:overlap}.
After time-stretching $\y_{\text{ref}}$ by a random value $t \sim \mathcal{U}(0.7, 1.5)$, the VQT representation is then randomly cropped to 252~bins and $w'=256$~time frames, i.e., 7 octaves and \SI{5.12}{\second}.

The encoder $\FF$ follows the ResNet-IBN architecture instantiated in~\citet{CLEWS}, and yields embedding vectors of dimensionality $m=2048$.
The temperature $\tau$ of our loss is initialized to 0.01 and trained in log-scale to prevent negative values.
We train the model using AdamW with an initial learning rate of $1.5 \cdot 10^{-3}$, which is divided by 5 when the training loss does not decrease during 5\,k steps. The batch size is $N = 384$, leading to a GPU memory usage of about \SI{75}{\giga\byte}.
We train our model on a single H100 GPU.
We refer to our code for the detailed configuration.

\section{Experiments}

\subsection{Setup}
\label{sec:setup}

Following previous works~\citep{van_balen_sample_2013,Cheston2025,Bhattacharjee2025}, we model automatic sample identification as a query-reference retrieval task.
We report the performance of our model and previous approaches using the Sample100 dataset~\citep{van_balen_sample_2013}, which consists of a set of 75~queries $\mathcal{Q}$ and a set of 68~reference songs $\mathcal{R}$, such that each track from $\mathcal{Q}$ samples a track from $\mathcal{R}$. In addition, a set $\mathcal{N}$ of 320~noise songs (i.e.,~songs not sampled by any of the queries) is incorporated. Both $\mathcal{Q}$, $\mathcal{R}$, and $\mathcal{N}$ are composed of commercial music recordings, $\mathcal{Q}$ being mostly hip-hop songs.

We also evaluate our model on a novel, manually curated dataset that we name SamplePairs. We release this dataset along with our code. It consists of 100 distinct pairs of commercial music, spanning various genres (pop, electro, hip-hop, etc.), plus 5,434 songs from MTG-Jamendo~\citep{MTGJamendo} to measure the ability of our model to retrieve candidate songs from a larger database. Here, query/candidate pairs are one-to-one mappings.


The evaluation considers all audio pairs $(\qq, \rr) \in \mathcal{Q} \times (\mathcal{R} \cup \mathcal{N})$. We split both $\qq$ and $\rr$ in overlapping chunks of \SI{5}{\second} with a hop size $h$, compute the pairwise cosine similarities between all chunks from $\qq$ and $\rr$, and finally report the maximal pairwise cosine similarity $s_{\qq \rr}$. This corresponds to a full-track best-match retrieval strategy. 
For each query $\qq$, we sort all reference and noise songs by decreasing $s_{\qq \rr}$ and measure the retrieval performance of our model by computing its mean average precision (mAP).
We also report the hit rate at $k$ (HR@$k$), that is, the proportion of queries for which the ground truth is in the top-$k$ closest reference tracks, as well as the mean and median normalized rank (mNR and medNR, respectively) as defined in \citet{SampleMatch,CLEWS}.
Since we observe that the hop size $h$ between test-time chunks affects the evaluation metrics, we report the metrics of the model that reaches the highest mAP among $h \in \{0.5, 1, 1.5, \dots 5\}$\,s, and further study the influence of $h$ in Sec.~\ref{sec:overlap}.

\subsection{Performance comparison}

\begin{table}[]
    \resizebox{\columnwidth}{!}{%
    \centering
    \begin{tabular}{lccc}
        \toprule
        Model & mAP & HR@1 & HR@10 \\
        \midrule
        Cheston et al.~\cite{Cheston2025} &
        0.441$^\dagger$ & - & - \\
        Bhattacharjee et al.~\cite{Bhattacharjee2025} &
        0.442$^\dagger$ & 0.155$^\dagger$ & 0.191$^\dagger$ \\
        Ours &
        0.603\ci{0.098} & 0.587\ci{0.111} & 0.733\ci{0.100} \\
        Ours + Top-5 retrieval &
        \textbf{0.622}\ci{0.099} & \textbf{0.600}\ci{0.110} & \textbf{0.747}\ci{0.098} \\
        \bottomrule
    \end{tabular}}
    \caption{Performances of our model against state-of-the-art baselines on the Sample100 dataset. $^\dagger$Results reported in \citet{Cheston2025,Bhattacharjee2025}.}
    \label{tab:sota}
\end{table}


Our results in Table~\ref{tab:baselines} highlight the relevance of our design choices.
In particular, we observe that time-stretching and pitch-shifting, which are typical transforms applied by artists while sampling, are essential for reaching good performances, as removing one of them substantially degrades the results.
Conversely, time-shifting seems moderately useful, particularly on Sample100 (we investigate its effect more in depth in Sec.~\ref{sec:overlap}).
Finally, to assess the relevance of our contrastive pipeline based on artificial mixes, we evaluate our method against a vanilla contrastive learning approach, in which the positive pairs are just the $(\yt_A^{(i)}, \yref^{(i)})$ and the loss is a regular NT-Xent contrastive loss~\citep{SimCLR}. This configuration also achieves a lower performance, but within the 95\,\% confidence interval.

Overall, we observe very large confidence intervals for all metrics, which can be attributed to the small size of both evaluation sets (75 and 100~queries, respectively) and to the presence of outliers (i.e.,~queries that our model fails to map close to their corresponding reference). This second hypothesis is supported by the important gap between the mNR and medNR. 
Finally, we note that the mNR and medNR metrics, which tend to be invariant to the number of reference/noise songs, are close between the two evaluation datasets, suggesting that the genre distribution (hip hop vs. diverse) has minimal impact on the quality of our model.

In Table~\ref{tab:sota}, we compare the performance of our model with state-of-the-art baselines on the Sample100 dataset. We observe that our method significantly outperforms existing ones. Beyond that, we propose a variant of our retrieval protocol, in which $s_{\qq\rr}$ is not the maximal pairwise similarity between chunks of $\qq, \rr$, but the average of the top-k similarities. The intuition is that sampled materials are usually looped and repeated multiple times throughout a song, especially in hip-hop. We observe a slight performance boost with this strategy and report the metrics for the top-5 version.

\subsection{Influence of hop size and frontend}
\label{sec:overlap}

\begin{figure}
    \centering
    \includegraphics[width=\linewidth]{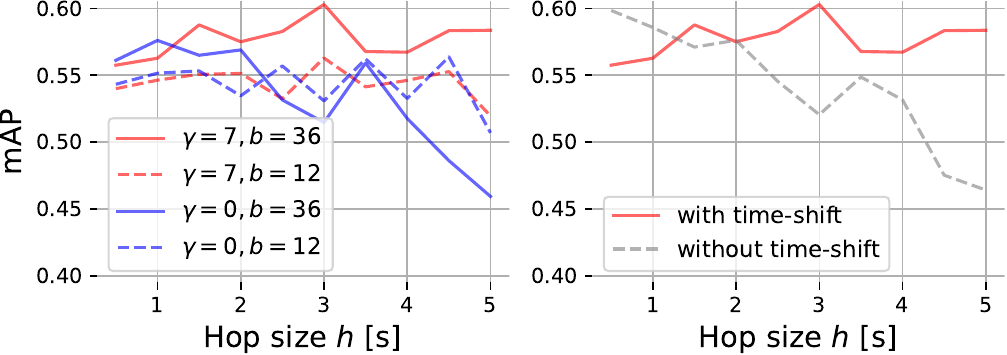}
    \caption{Influence of $h$ on the mAP on Sample100, for different frontends (left) and with or without time-shift (right).}
    \label{fig:overlap}
\end{figure}

We now study mAP as a function of the hop size $h$ when using a VQT (with bandwidth parameter $\gamma = 7$) and a regular CQT ($\gamma = 0$), both with $b=12$ or $b=36$~bins per octave (Fig.~\ref{fig:overlap}, left).
Interestingly, when keeping a low resolution ($b = 12$), the frontend and the hop size have almost no influence. However, when the resolution of the CQT increases to $b=36$, the model achieves a better performance when $h$ is small, but it degrades significantly with longer hop sizes. Using a VQT with increased $\gamma=7$ circumvents this issue and leads to maximal performances that remain stable across different overlaps. Finally, we observe that removing time-shifting, which seemed harmless in Table~\ref{tab:baselines}, actually hurts performance when $h$ increases (Fig.~\ref{fig:overlap}, right). This is intuitive, as time-shifting provides the model with positive pairs that do not fully overlap, thereby increasing its robustness to time misalignment.

\subsection{Robustness to the presence of noise songs}

\begin{figure}
    \centering
    \includegraphics[width=\linewidth]{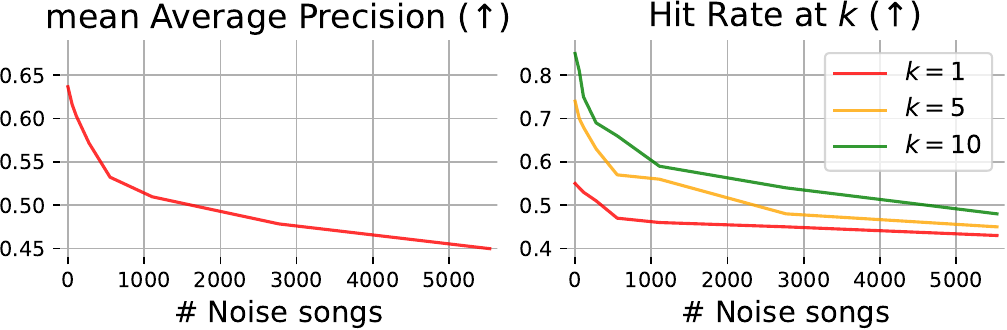}
    \caption{Influence of the number of noise songs in the SamplePairs dataset on the evaluation metrics.}
    \label{fig:noise}
\end{figure}

In real-world scenarios, there is no fixed reference set; instead, for each query, we aim to identify the correct reference within a very large catalog. To move towards this more realistic scenario, 
we study how progressively adding noise songs in the SamplePairs dataset affects the retrieval performance (Fig.~\ref{fig:noise}). We find that adding a few noise songs leads to a large drop in performance, whereas increasing the amount further only results in a much smaller gap. In particular, HR@1 remains almost unchanged even with a large amount of noise songs, and the very small difference between HR@5 and HR@10 suggests that the learned representations are highly clustered. We also observed (not shown here) that the normalized rank remains almost constant with respect to the amount of noise songs.


\subsection{Influence of the training dataset}

\begin{figure}
    \centering
    \includegraphics[width=\linewidth]{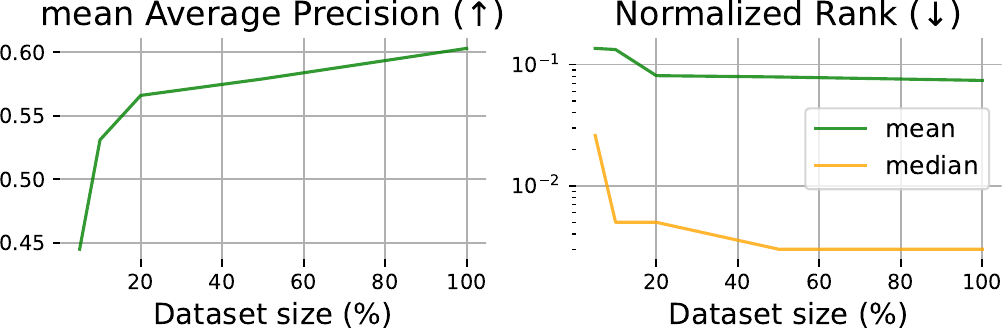}
    \caption{Influence of the train set size on the final performances of our model, measured on Sample100.}
    \label{fig:data}
\end{figure}

Even if we empirically validated our design choices in Table~\ref{tab:baselines}, we acknowledge that the main gap between our model and existing baselines, illustrated in Table~\ref{tab:sota}, can be attributed to our large-scale multi-track training set.
In this section, we therefore measure which aspects of this dataset contribute the most to the reported performance. We specifically study the influence of two parameters: the size of the dataset and the availability of ground truth stems.
First, we report the metrics of our model when training it on a random subset of our data. Our results, depicted in Figure~\ref{fig:data}, reveal that the effect of the dataset size is limited.
Although metrics improve with more data, training with only 20\,\% of it ($\approx$4\,k songs) already yields performances close to those achieved with the full dataset. More surprisingly, training with only 5\% of the data ($\approx$1,000 songs) already yields a mAP that is comparable to previous state-of-the-art baselines, which are trained on larger datasets.

Another benefit of our dataset compared to previous approaches is the availability of fine-grained ground truth separated instruments. To study this effect, we retrain our model on a modified version of our dataset in which we merge instruments into 4 or 6 stem categories (bass, drums, vocals, and other, plus optionally piano and guitar). In addition, we also measure the effect of using source separation when ground truth stems are not available. To do so, we mix all instruments of each song together and retrieve 4 stems using the open state-of-the-art HT-Demucs~\citep{HTDemucs}, which was used in \citet{Cheston2025,Bhattacharjee2025} to create their artificial dataset. Our results, reported in Table~\ref{tab:sources}, highlight that reducing the number of sources per song degrades all metrics, with reducing to 6 stems having as much effect on mAP as using only 20\% of the dataset (see Figure~\ref{fig:data}). In addition, relying on source separation significantly hurts performance, even compared to using 4 ground truth stems. Overall, our findings suggest that having high-quality ground truth separated stems yields more improvement than simply scaling up data.

\begin{table}[]
    \resizebox{\columnwidth}{!}{%
    \centering
    \begin{tabular}{lcccc}
        \toprule
        Model & mAP ($\uparrow$) & HR@1 ($\uparrow$) & 
        mNR ($\downarrow$) & medNR ($\downarrow$) \\
        \midrule
        Ours &
        \textbf{0.603}\ci{0.098} & \textbf{0.587}\ci{0.111} &   \textbf{0.074}\ci{0.036} & \textbf{0.003} \\
        6 stems & 
        0.557\ci{0.102}  & 0.560\ci{0.112} &  
        0.085\ci{0.036} & \textbf{0.003} \\
        4 stems &
        0.527\ci{0.101} & 0.520\ci{0.113} &  
        0.083\ci{0.038} & 0.008 \\
        Demucs &
        0.466\ci{0.103} & 0.453\ci{0.113} &  
        0.130\ci{0.049} & 0.026 \\
        \bottomrule
    \end{tabular}}
    \caption{Influence of the number of stems used during training on our performances on the Sample100 dataset.}
    \label{tab:sources}
\end{table}

\section{Conclusion}

In this paper, we propose a novel method for automatic music sample identification.
We leverage a dataset of multi-track recordings to create artificial mixes on-the-fly and propose a novel contrastive objective to train our model, achieving state-of-the-art performance and outperforming previous methods by a significant margin.
Our experiments validate the relevance of our design choices and highlight, in particular, the need for high-quality stem data.

However, despite its success, our method still suffers from a conceptual limitation: if B and C sample two different instruments from A, an ideal system should map A close to B and C, but B and C should still be far from each other, which would contradict the triangular inequality and the typical contrastive loss formulations. 
More generally, the fundamental limitations of single-embedding models for retrieval applications have recently been studied in \citet{Weller2025}.
Future work on automatic sample identification should therefore deal with these issues while remaining tractable for large-scale applications.

\section{Acknowledgments}

We would like to thank Fabio Morreale, Eleonora Mancini and Dipam Goswami for insightful discussions throughout this project.

\bibliographystyle{ieee}
\bibliography{refs}

\end{document}